\documentstyle[preprint,prd,aps]{revtex}
\global\arraycolsep=2pt
\makeatletter
\def\fmslash{\@ifnextchar[{\fmsl@sh}{\fmsl@sh[0mu]}}
\def\fmsl@sh[#1]#2{  \mathchoice
    {\@fmsl@sh\displaystyle{#1}{#2}}    {\@fmsl@sh\textstyle{#1}{#2}}    
{\@fmsl@sh\scriptstyle{#1}{#2}}    {\@fmsl@sh\scriptscriptstyle{#1}{#2}}}
\def\@fmsl@sh#1#2#3{\m@th\ooalign{$\hfil#1\mkern#2/\hfil$\crcr$#1#3$}}
\makeatother

\begin{document}
\title{Reply to Comment}
\author{Ping-Xing Chen and Lin-Mei Liang}
\address{Laboratory of Quantum Communication and Quantum Computation, \\
University of Science and Technology of
China, Hefei, 230026, P. R. China \\
and \\
\thanks{%
Corresponding address}Department of Applied Physics,\\ National University of
Defense Technology,
Changsha, 410073, 
P. R. China}
\author{Cheng-Zu Li}
\address{Department of Applied Physics,\\
National University of Defense Technology,
Changsha, 410073, P. R. China }
\author{Ming-Qiu Huang}
\address{CCAST (World Laboratory) P.O. Box 8730, Beijing, 100080, China\\
and Department of Applied Physics,\\
National University of Defense Technology,
Changsha, 410073, P. R. China }
\date{\today }
\maketitle

\thispagestyle{empty}

\pagenumbering{arabic} 

In the comment \cite{4} Eggeling, Vollbrecht and Wolf suspect our method in Ref. \cite
{1} is not practical. Here we explain our result and method and show that
our example can tell one how to judge a separable state, and so our method
is practical, at least for many mixed states.

$\ \qquad \qquad \qquad \qquad \qquad \qquad \qquad ${\bf A. Explanation}

Recently we considered the necessary and sufficient condition of
separability and gained a practical criterion \cite{1}. Our main idea is as
followings:

1. We decompose a mixed state $\rho $ of $m\otimes n$ dimension Hilbert
space into (m-1)(n-1) pairs which are in 2$\otimes 2$ dimension Hilbert
space, from Ref.\cite{2}, we get the necessary condition of each pair being
parallel, i.e. $\rho $ being separable.

2. Each parallel pair corresponding to a set of matrices U, if there are
intersection of all matrices U, $\rho $ is separable, otherwise $\rho $ is
inseparable. We shown by our example\cite{1} that judging the intersection
existing or not comes down to judging a set of special symmetric quadratic
equations having solutions or not, which is not difficult, so our method is
practical.

$\ \qquad \qquad \qquad \qquad \qquad \qquad \qquad ${\bf B. Reply}

The first step above generalizes Wootter's work\cite{2} to higher dimension.
This is meaningful not only for separable criterion but also for other
approach(we are working in this). Parallel pair is different from PPT\cite{3}%
, though they are equivalent for 2$\otimes 2$ system, but in the case of
higher dimension they have different character and usage.

If one has found a matrix U which brings up a set of separable pure states
decomposition of $\rho ,$one can say $\rho $ is separable. But how to judge
this matrix existing or not? The second step above has told one how to find
this matrix U, so it is practical.

We think if the authors of the comment\cite{4} consider the example in Ref%
\cite{1} carefully, they will find our method is practical.


\begin{references}
\bibitem{4}  T.Eggeling, K.G.H.Vollbrecht, M.M.Wolf quant-ph/0103003

\bibitem{1}  P.X.Chen, Lin-Mei Liang, Cheng-Zu Li and Ming-Qiu Huang,
quant-ph/0102133,Phys.Rev.A.(in press)

\bibitem{2}  W. K.Wootters, Phys. Rev. Lett. {\bf 80}, 2245(1998).

\bibitem{3}  A. Peres, Phys. Rev. Lett. {\bf 77}, 1413(1996).
\end{references}
\end{document}